\documentclass[a4paper,fleqn,usenatbib]{mnras}

\usepackage[T1]{fontenc}
\usepackage{ae,aecompl}

\usepackage{graphicx}	
\usepackage{amsmath}	
\usepackage{amssymb}	

\newcommand{\param}[1]{\mbox{\texttt{#1}}}

\title[Edge-on galaxies in the HST COSMOS field]{Edge-on galaxies in the HST COSMOS field:
the evolution of stellar discs up to z$\sim$0.5}

\author[P.A. Usachev, V.P. Reshetnikov, S.S. Savchenko]{
	Pavel A. Usachev$^{1,2}$, Vladimir P. Reshetnikov$^{1,2}$\thanks{E-mail: v.reshetnikov@spbu.ru}, 
	Sergey S. Savchenko$^{1,2}$
	\\
$^1$ St.Petersburg State University, 7/9 Universitetskaya nab., St.Petersburg, 
199034 Russia \\
$^2$ Special Astrophysical Observatory, Russian Academy of Sciences, 
Nizhnii Arkhyz, 369167 Russia
}

\date{Accepted 2023. Received 2023; in original form 2023}

\pubyear{2022}

\begin{document}
	\label{firstpage}
	\pagerange{\pageref{firstpage}--\pageref{lastpage}}
	\maketitle
	
\begin{abstract}

We present a sample of 950 edge-on spiral galaxies found with the use of 
an artificial neural network in the Hubble Space Telescope COSMOS field. 
This is currently the largest sample of distant edge-on galaxies.
For all galaxies we analyzed the 2D brightness distributions in the F814W filter and
measured the radial and vertical exponential scales ($h$ and $h_z$ correspondingly)
of the brightness distribution. By comparing the characteristics of distant galaxies 
with those of nearby objects, we conclude that thin stellar discs with $h/h_z \geq 10$ 
at $z \approx 0.5$ should be rarer than today. Both exponential scales of the stellar 
disc show evidence of luminosity-dependent evolution: in faint galaxies the 
$h$ and $h_z$ values do not change with $z$, in bright (and massive) spiral galaxies 
both scales, on average, grow towards our epoch.

\end{abstract}




\section{Introduction}

Galaxies evolve --  their mass, size, morphological type, bulge and disc parameters, 
and all other possible characteristics change over time (e.g. \citealt{con2014}). 
Observations of galaxies at different 
redshifts make it possible to study this evolution directly. At present, a huge amount 
of observational data on galaxies in the surrounding Universe has been accumulated 
and, thanks to the work of space observatories and large ground-based telescopes, 
information on distant galaxies is also gradually increasing. 

The study of edge-on spiral galaxies is a very important area of extragalactic research.
Such orientation of galaxies provides a unique opportunity to simultaneously study both 
the radial and vertical structure of their stellar discs (e.g.
\citealt{burstein1979, hk1979, vdks1981, bd1994, dg1998, EGIS}). 
Also, the study of such objects allows us to investigate the distribution and properties 
of dust in the discs of galaxies (e.g. \citealt{xil1998, deg2014, mosu2022}), 
the characteristics of their dark halos (\citealt{zasov1991, ob2010, biz2021}), 
shape of galactic discs and bulges (e.g. \citealt{msr2010, rsm2015, EGIPS, marchuk2022}), 
properties of warps (\citealt{rmm2016}) and many other issues.

At the moment, much material has been collected on the structure of nearby edge-on galaxies, 
while the photometric structure of distant spirals has not yet been sufficiently 
studied. Previous studies were based on small samples of galaxies (a few dozens of objects) 
observed in the Hubble Space Telescope (HST) deep fields, such as HDF-N, HDF-S, HUDF etc.
In these studies it was shown that the relative thickness (ratio of vertical to radial 
scales) of stellar discs at redshift $z \sim 0.5-1$ exceeds, on average, 
the thickness of the discs of nearby spiral galaxies (\citealt{rdc2003, elm2005, elm2006}).
More recent studies have shown that the vertical scales of the surface brightness distribution 
of the discs of spiral galaxies do not show appreciable changes at $z \leq 1$ and therefore 
the excess relative thickness of distant discs is most likely due to the evolution of their 
radial scales (\citealt{rus2019, ru2021}). 

The main goals of this work are to compile a large sample (an order of magnitude larger 
than in previous works) of edge-on spiral galaxies at $z \sim 0.5$ and to analyse the 
brightness distribution in these galaxies. In this paper we describe this sample, 
analyse the two-dimensional brightness distributions in the sample galaxies, and investigate 
the possible evolution of the stellar discs characteristics with redshift. 

Throughout this work, we adopt a standard flat $\Lambda$CDM
cosmology with $\Omega_m$=0.3, $\Omega_{\Lambda}$=0.7, 
$H_0$=70 km\,s$^{-1}$\,Mpc$^{-1}$. All magnitudes in the paper are given
in the AB-system.

\section{The sample of edge-on galaxies}

As a source of data on edge-on galaxies, we considered the COSMOS HST ACS field in the F814W filter \citep{COSMOS1,
  COSMOS2}. This field is $\sim 1.3^{\circ} \times 1.3^{\circ}$ in size and it contains hundreds of thousands of
relatively bright galaxies up to redshifts $z > 5$ (e.g. \citealt{COSMOS2020}). For the purposes of selection, we used a
neural network, that was developed and trained to search for edge-on spiral galaxies (see \citealt{EGIPS, sav2023} for
details of the network architecture and the training process). Processed images of 26\,113 galaxies with
$I\,(\text{F814W}) < 22\fm5$ from \cite{stamps} as well as weight maps and PSF data for each galaxy were taken as the
initial data.

The mentioned neural network has been adapted and has undergone special training for working with prepared images of
COSMOS galaxies in order to obtain the best results when taking into account the technical features of COSMOS field
images: their angular resolution (pixel size is 0.$''$03), average S/N ratio, degree of crowding, and, most importantly,
the presence of data in only one colour filter (F814W).

The algorithm for selecting for edge-on galaxies consisted of two stages. First, with the help of the neural network,
2\,586 candidates for edge-on galaxies were obtained with a relative confidence degree of the neural network in the
answer of at least 95\%.  Second, the images of all candidates were reviewed by the authors of this paper in order to
select only those objects that are relatively confidently visually identified as being seen edge-on. A final sample of
950 edge-on galaxies in the COSMOS field was thus compiled. It is worth noting that the sample is biased towards
normal-looking (symmetrical, centrally-concentrated) galaxies, and it does not include 
pecular objects (e.g. with a strong warp).

The sample galaxies were identified with the COSMOS2020 catalogue (\citealt{COSMOS2020}).  For each object we took the
photometric redshift found with the LePhare code (\citealt{ilbert2006}), the rest-frame absolute magnitude in the $R$
band (Subaru HSC), and the stellar mass. If the LePhare results were not available, we used the redshifts found with the
EAZY package (\citealt{EAZY}) or took redshifts from catalogue by \cite{ilbert2009}. In total, we found characteristics
for 941 edge-on galaxies.

\begin{figure}
\centering
\includegraphics[width=4.0cm, angle=-90, clip=]{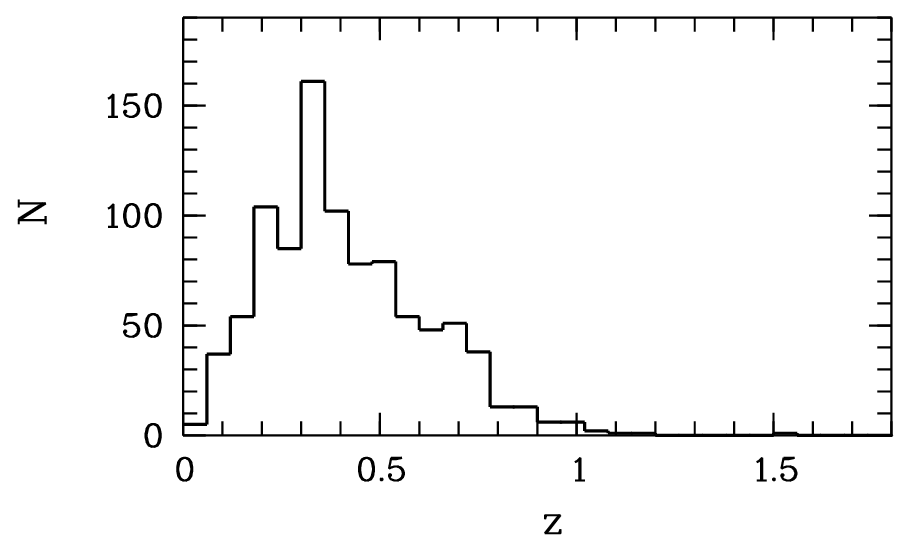}
\includegraphics[width=4.0cm, angle=-90, clip=]{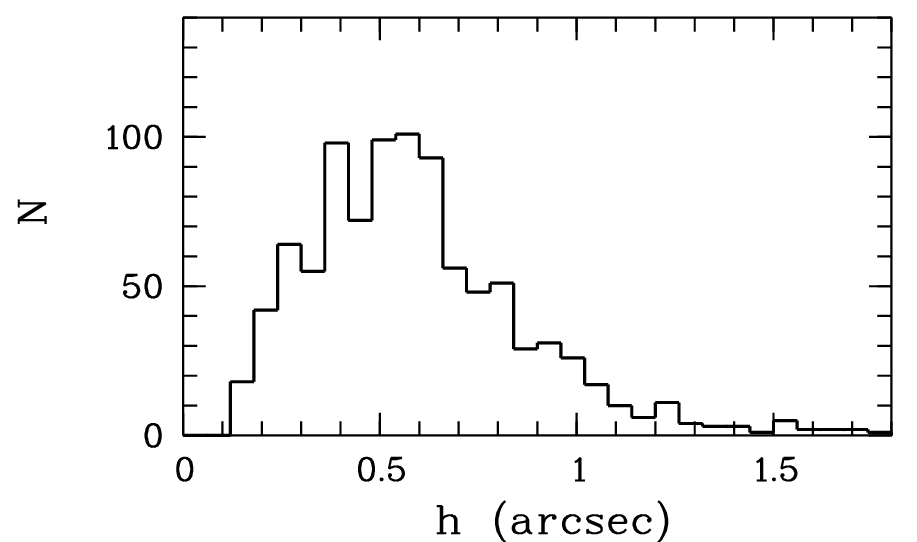}
\includegraphics[width=4.0cm, angle=-90, clip=]{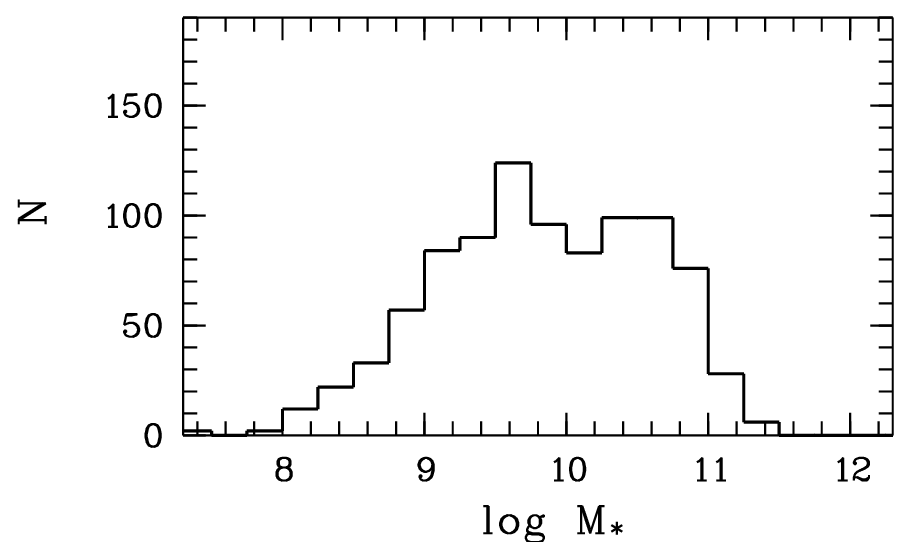}
\caption{Distributions of redshift (top), radial scalelength (middle), 
and stellar mass in solar units (bottom) for the galaxies in our sample. }
\label{fig:sample}
\end{figure}

\autoref{fig:sample} illustrates some general features of the sample: distributions of galaxies by redshift, exponential
scalelength (see Sect.~3) and stellar mass. As can be seen in the figure, the sample includes galaxies up to
$z \approx 1$ with the mean redshift of 0.42$\pm$0.24. The galaxies are small in angular size: mean observed value of
exponential scalelength of their stellar discs is 0$\farcs$59$\pm$0$\farcs$28. Edge-on galaxies in the COSMOS field span
a stellar mass range $8\,\leq\,$\,log\,M$_*$/M$_{\odot}\,\leq$\,11.5 with the mean value of 9.82$\pm$0.78.

\section{Two-dimensional decomposition}

\begin{figure}
\centering
\includegraphics[width=8.3cm]{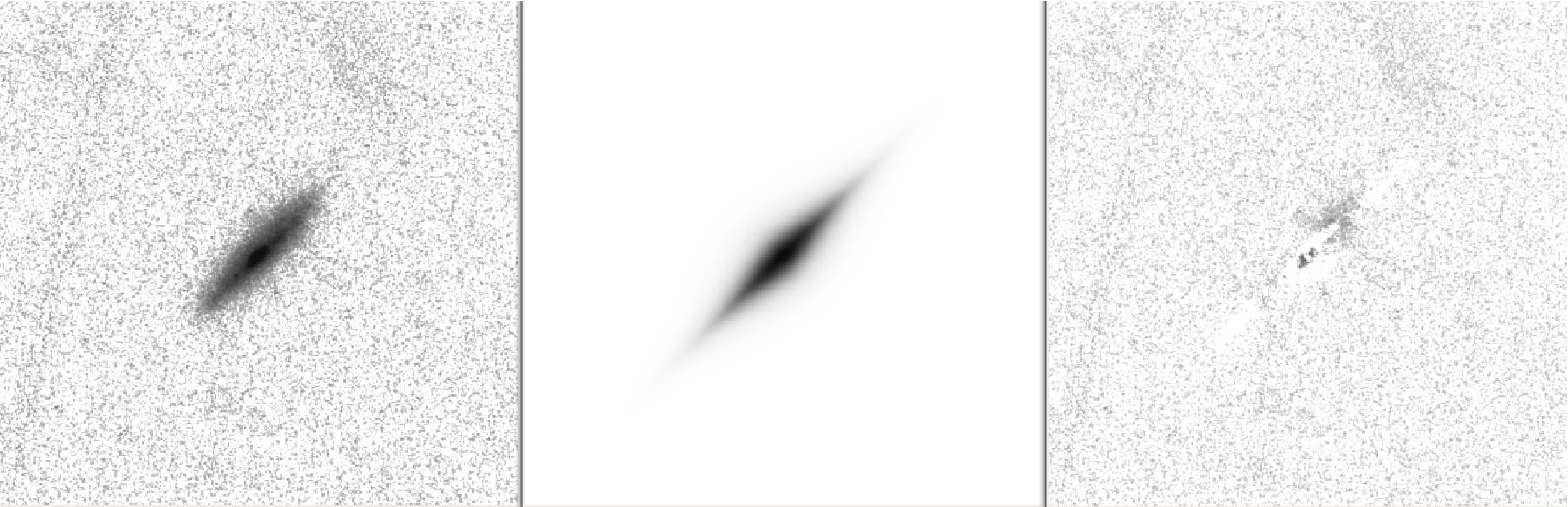}
\includegraphics[width=8.3cm]{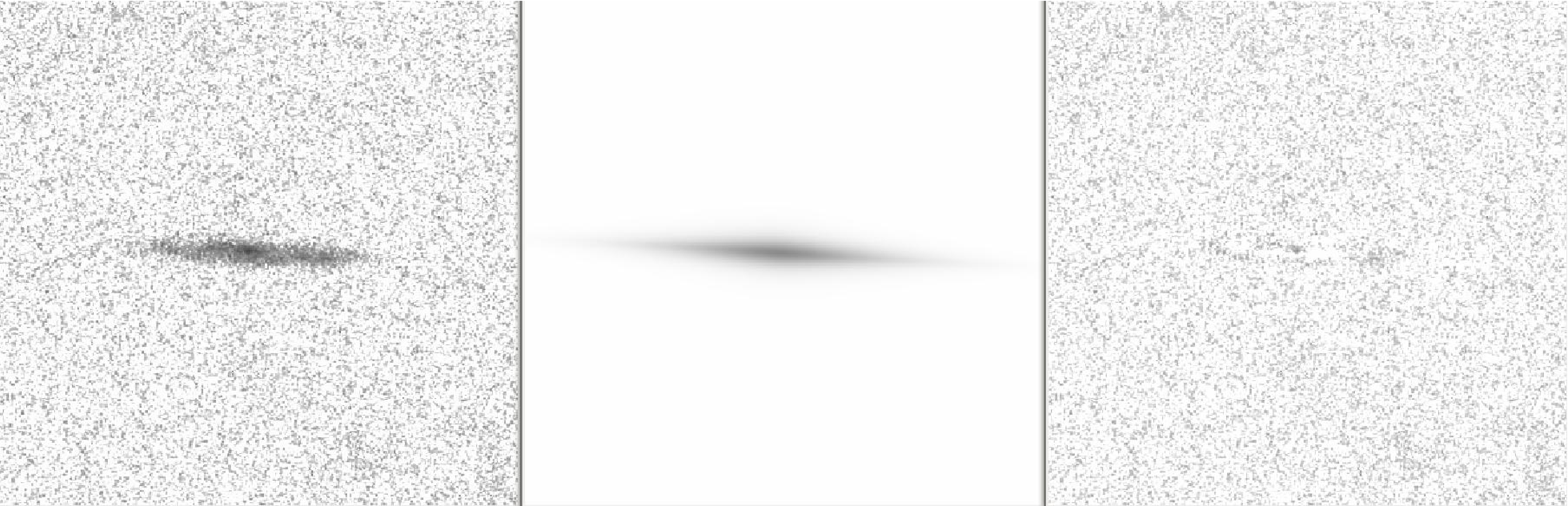}
\includegraphics[width=8.3cm]{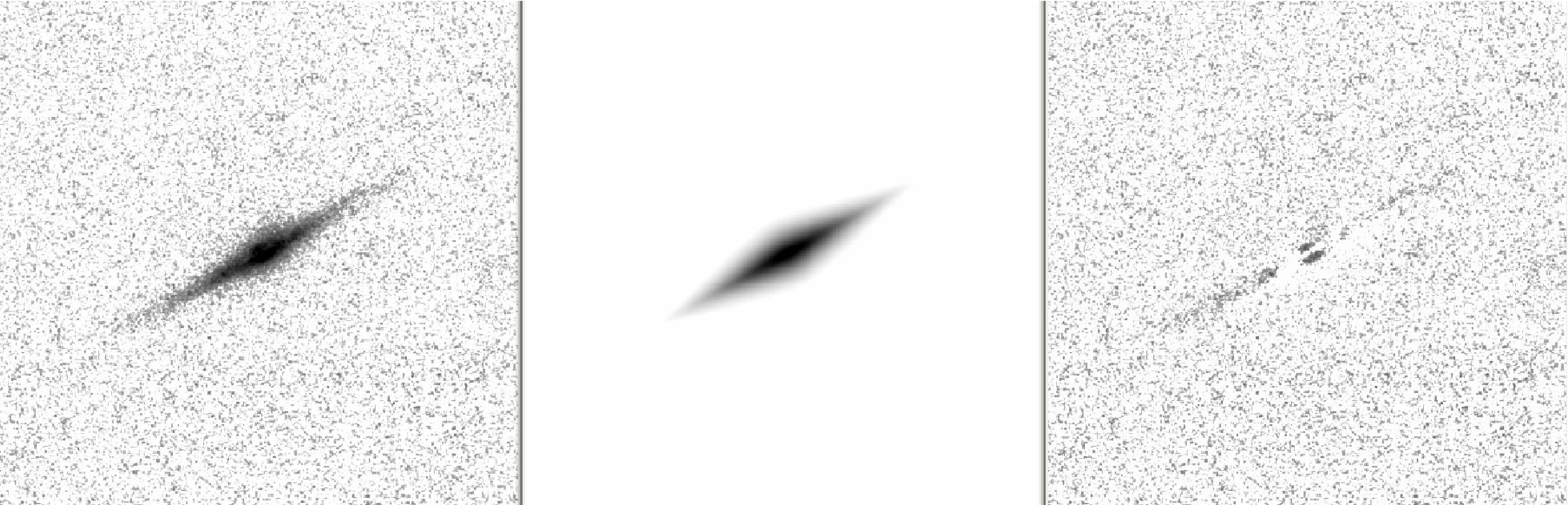}
\caption{Decomposition results for three randomly selected galaxies. Left panels show
original images, middle panels present the models, right panels give the residual images.
Image sizes are 12$\arcsec$, 17$\arcsec$, 12$\arcsec$ (from top to bottom).}
\label{fig:examples}
\end{figure}   

To describe the photometric structure of distant edge-on galaxies, we chose the simplest two-component model: 1) a disc
in which the brightness distribution in radial and vertical (perpendicular to the disc plane) directions is described by
an exponential law with corresponding scales $h$ and $h_z$, and 2) bulge described by a symmetric two-dimensional
Gaussian or a point source following the PSF form when the central component is not resolved. Both components share the
same center in the image.  Hereafter we will discuss the characteristics of galactic discs only.

Images of all 950 galaxies were decomposed with the \textsc{imfit} program 
(\citealt{imfit}) using convolution with PSF and with weight images provided along with 
the original galaxy images in the COSMOS field (\citealt{stamps}). 
We used the \textsc{EdgeOnDisk} function in the \textsc{imfit} package 
to describe the disc. To set the exponential distribution in the vertical direction 
in the \textsc{EdgeOnDisk} function, we have accepted the parameter $\param{n}=1\texttt{e}6$ --
see formula (36) in \citet{imfit}.

Several examples of our photometric modeling are presented in
\autoref{fig:examples}.

\section{Results and discussion}

\subsection{Scales of distant discs}

\begin{figure}
\centering
\includegraphics[width=14cm, angle=-90, clip=]{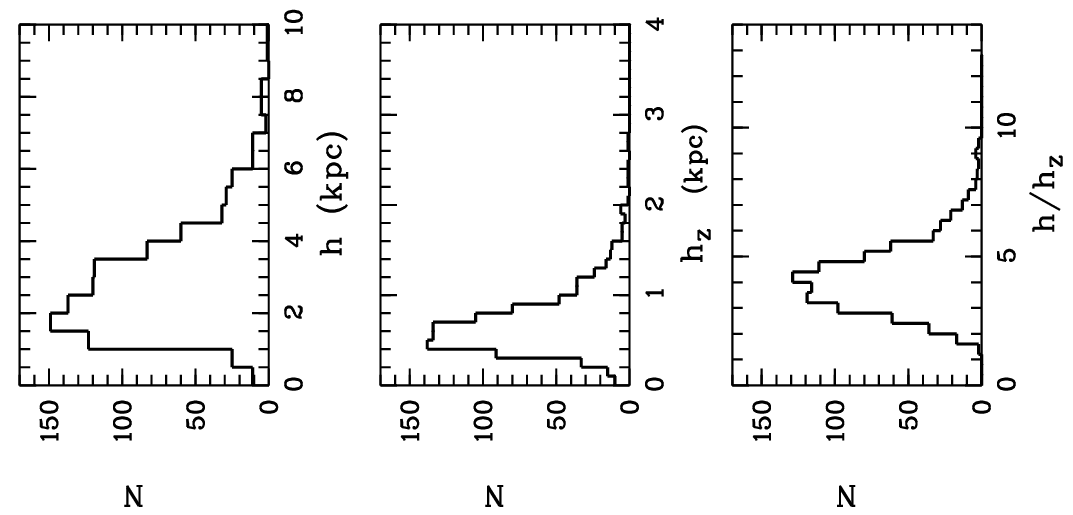}
\caption{Distributions of the sample galaxies in exponential scalelength $h$ (top),
exponential scaleheight $h_z$ (middle), and the $h/h_z$ ratio (bottom).}
\label{fig:hist}
\end{figure}

The mean values of the exponential scales for the edge-on galaxies in the COSMOS field
are $\langle h \rangle  = 2.91 \pm 1.49$ kpc and $\langle h_z \rangle  = 0.72 \pm 0.36$ kpc
(see also \autoref{fig:hist}). The corresponding median values are 2.64 kpc and 0.64 kpc.
Both values are typical for the non-dwarf spirals in the nearby Universe 
(e.g., \citealt{EGIS}). It should be noted that we cannot separate the contributions 
of thin and thick discs in distant galaxies and so we measure the composite thickness 
of these discs. When comparing with galaxies at $z \approx 0$, we will also use the
composite thicknesses of nearby discs.

The mean value of the scales ratio is $\langle h/h_z \rangle = 4.20 \pm 1.30$
(median is 4.08) (\autoref{fig:hist}). This value is close to that found for galaxies in the HUDF 
but, on the other hand, it looks smaller (i.e., the galactic disks are thicker) than those for 
spiral galaxies in the local Universe -- see sect.~3.3 in \citet{rus2019}.
Here and in the following, we assume that our sample of distant edge-on galaxies 
and the nearby galaxies have a similar distribution in inclination
and that there is no dependence of galaxy inclination on redshift in the COSMOS sample.

\begin{figure}
\centering
\includegraphics[width=6cm, angle=-90, clip=]{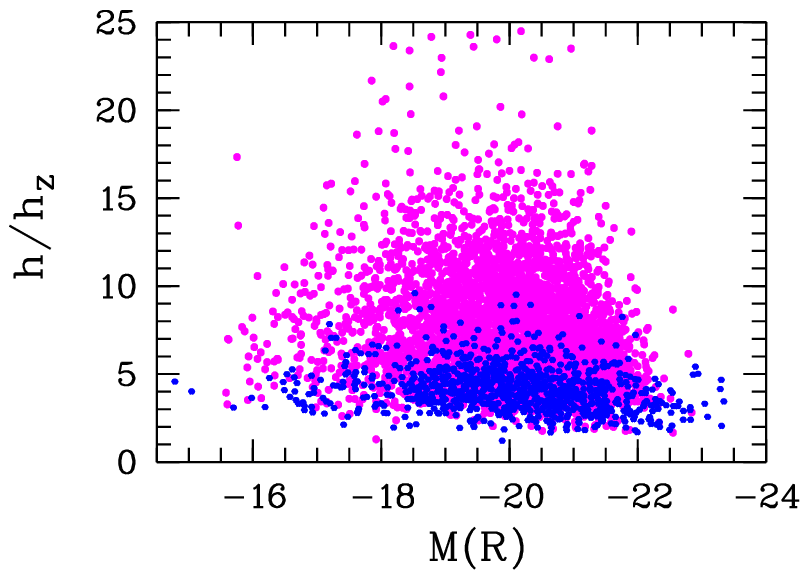}
\caption{Distribution of the galaxies from the HST COSMOS field on the galaxy absolute 
magnitude $M(R)$ -- scalelength to scaleheight ratio plane in the F814W filter
(blue dots). The magenta dots represent the characteristics of nearby spiral galaxies 
from the SDSS in the $r$ band (\citealt{EGIS}).}
\label{fig:absscales}
\end{figure}   

\autoref{fig:absscales} presents the positions of our edge-on galaxies on the
absolute magnitude in the $R$ filter ($M(R)$) -- $h/h_z$ plane. We also plotted
the galaxy characteristics from the SDSS survey in the $r$ filter according 
to \citet{EGIS} onto this plane. When comparing the results, we assume
$h_z = z_0/2$, where $z_0$ is the thickness parameter used in \citet{EGIS}.
\autoref{fig:absscales} shows that nearby galaxies 
from the SDSS survey are very widely spaced on this plane -- many galaxies with 
$h/h_z \geq 10$ and even with $h/h_z \geq 15$ are observed. On the other hand, the 
distant galaxies show a much smaller spread of $h/h_z$ values and, in addition, over 
the entire luminosity range they are located mainly along the lower envelope of the 
distribution of close galaxies. If this feature is not explained by observational
selection (i.e., the difficulty in detecting very thin discs as $z$ increases), it
could mean that thin stellar discs are apparently rare at $z \sim 0.5$.
If confirmed, the conclusion about the rarity of thin stellar discs at $z \sim 0.5$ 
may mean that thin stellar discs of galaxies surrounding us were formed mainly in the 
last 4--5 Gys.

\autoref{fig:absscales} also illustrates some tendency for brighter (and more massive)
galaxies to have relatively thicker stellar discs. This trend is seen for distant 
galaxies and is traced by the lower envelope of the distribution for nearby galaxies.

\subsection{Evolution of the discs}

\begin{figure}
\centering
\includegraphics[width=12.5cm, angle=-90, clip=]{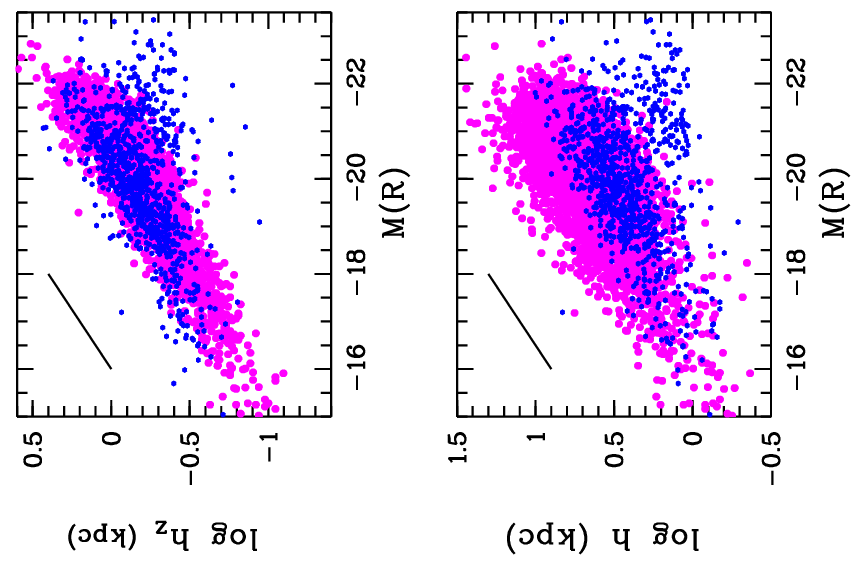}
\caption{Distribution of the galaxies from the COSMOS field (blue dots) 
on the $M(R)$ -- disk scaleheight (top) and $M(R)$ -- disk scalelength (bottom) planes. 
The magenta dots indicate the characteristics of nearby 
spiral galaxies in the $r$ band from \citet{EGIS}. Straight line segments 
illustrate the slope of the line of constant luminosity for a purely exponential disc.}
\label{fig:abshhz}
\end{figure}   

We compare the characteristics of the galaxies from the COSMOS field with
the parameters of relatively nearby objects on the galaxy absolute magnitude --
vertical or radial scale in \autoref{fig:abshhz}.
As can be seen in the figure, the vertical scales ($h_z$) of most distant galaxies are 
within the region occupied by nearby objects. Some of bright galaxies ($M(R) \leq -20^m$)
deviate from this region towars smaller $h_z$ values.

For the scalelengths $h$, the situation is similar (\autoref{fig:abshhz}) -- 
most of the distant galaxies are 
located within the region for nearby galaxies, though with a shift towards smaller values; 
among the bright objects ($M(R) \leq -20^m$), some are strongly shifted towards more 
compact discs.

It can be assumed that the differences between the distributions for nearby and distant
galaxies in \autoref{fig:abshhz} reflect the evolution of the structure of stellar discs.
In this case, the observed differences may indicate differential evolution of
$h$ and $h_z$ values. According to \autoref{fig:abshhz}, we can conclude that for
galaxies with $M(R) \geq -20^m$ (or M$_* \leq $\,10$^{10}$\,M$_{\odot}$), the scaleheight
shows no appreciable evolution to $z \sim 0.5$. For brighter and more massive galaxies,
one should allow a gradual thickening of the discs by the modern epoch (see also
\citealt{hc2023}).

The radial structure of discs indicates a more pronounced evolution. If, to a first
approximation, we assume that the scalelength changes with redshift as $(1 + z)^{-n}$, then our
data indicate a strong dependence of $n$ on the luminosity (mass). Thus, at $M(R) \geq -18^m$
$n \approx 0$ (no evolution), since the distant and nearby galaxies are in the same 
region (see \autoref{fig:abshhz}). In order to align the position of distant galaxies with
$-20^m \leq M(R) \leq -18^m$ with the middle of distribution of the SDSS galaxies
in \autoref{fig:abshhz}, it is necessary to allow the scalelength evolution with $n \sim 1$.
The brightest galaxies ($M(R) \leq -20^m$) require the most significant evolution 
with $n \sim 2$. Assuming a moderate evolution of galaxy luminosity 
($\Delta M \propto z\times(0.5-1)^m$), this does not noticeable change the
$n$ values.

In previous works, the half-light radius was usually used to study the variation
of galaxy size with redshift (e.g., \citealt{bouwens2004, shibuya2015}).
The results of these works are in qualitative agreement: the sizes of 
galaxies (star-forming and quenched) grow with time, and this growth depends on the mass -- 
more massive galaxies grow faster (e.g., \citealt{allen2017, kaw2021, ned2021, buit2023}).
In our work, using not the effective galaxy radius, but direct measurements of the 
scales of stellar discs up to $z \sim 0.5$, we confirmed the existence of the evolution 
of spiral galaxy sizes, as well as the dependence of the evolution on the luminosity (mass) of 
galaxies.

\section{Conclusions}

Using a neural network, we have compiled a sample of 950 edge-on galaxies
in the HST COSMOS field. This is currently the largest sample of distant
edge-on galaxies. The sample galaxies are at a mean redshift of 0.42 and 
are distributed by stellar mass in the range $8\,\leq\,$\,log\,M$_*$/M$_{\odot}\,\leq$\,11.5.
We have performed a two-dimensional decomposition of the galaxy images in the
F814W filter, determined the characteristics of their stellar discs, and compared them 
with those of galaxies at $z \approx 0$.

Our main conclusions are as follows:

-- The observed {\it relative} thickness of the stellar discs of distant galaxies 
apparently exceeds 
the thickness of the discs in the nearby Universe, but this result may be 
influenced by the existence of possible selection of distant galaxies by inclination angle.
Thin stellar discs with $h/h_z \ge 10$ appear to be less common at $z \approx 0.5$ than at
$z \approx 0$.

-- The vertical scale of disks of galaxies with $M(R) \geq -20^m$ 
(or M$_* \leq $\,10$^{10}$\,M$_{\odot}$) does not show notable evolution at $z \leq 0.5$.
Discs of more massive galaxies must on average thicken towards $z \approx 0$.
We emphasise that we discuss composite profiles in which the contributions 
of thick and thin stellar discs are not separated. The thick and thin discs may have 
different origins, chemical composition, colour indices, and other characteristics
(e.g. \citealt{kasp2016, kasp2020}). 
They may also show different redshift evolution and therefore their varying contributions 
at different $z$ can influence the composite photometric profile.
The study of the evolution of two different types of stellar discs requires further 
detailed investigation.

-- Exponential scalelength of the discs shows signs of differential evolution -- for low 
luminosity galaxies ($M(R) \geq -18^m$) it does not change with redshift, for bright 
galaxies it grows from $z \approx 0.5$ to current epoch.

We hope that current and future multicolour galaxy surveys (like JWST CEERS project) will develop 
and extend to earlier cosmological epochs the issues raised in our paper.

\section*{Acknowledgements}

We are thankful to the referee for useful comments.
This research was supported by the Russian Science Foundation
grant No 19-12-00145.

\section*{Data availability}

The data underlying this article will be shared on reasonable request to the 
corresponding author.



\bibliographystyle{mnras}
\bibliography{art}

\label{lastpage}
\end{document}